\documentclass[letterpaper,prl,twocolumn,nobalancelastpage]{revtex4}

\usepackage{graphicx}
\usepackage{amsmath}
\usepackage{amssymb}
\usepackage{units}
\usepackage{array}

\begin{document}

\title{Long range Coulomb interactions in bilayer graphene}

\author{D. S. L. Abergel}
\email{abergel@cc.umanitoba.ca}
\author{Tapash Chakraborty}
\affiliation{Department of Physics and Astronomy, University of
Manitoba, Winnipeg MB, R3T 2N2, Canada.}

\begin{abstract}
We report on our studies of interacting electrons in bilayer graphene in
a magnetic field.
We demonstrate that the long range Coulomb interactions between
electrons in this material are highly important and account for the
band asymmetry in recent optical magneto-absorption experiments
\cite{henriksen:prl100}.
We show that in the unbiased bilayer (where both layers are at the same
electrostatic potential), the interactions can cause mixing of Landau
levels in moderate magnetic fields. 
For the biased bilayer (when the two layers are at different
potentials), we demonstrate that the interactions are responsible for a
change in the total spin of the ground state for half-filled Landau
levels in the valence band.
\end{abstract}

\maketitle

Monolayer graphene is a two-dimensional hexagonal crystal of carbon
atoms whose gapless, relativistic-like, linear low energy dispersion has
made it the subject of intense study since it was first isolated in 2004
\cite{novoselov:sci306}. 
Bilayer graphene (BLG) \cite{mccann:ssc143}, the subject of our present
study consists of a pair of monolayers bound by relatively weak dimer
bonds formed perpendicular to the monolayer planes.
Both the conduction and valence bands have low energy structure
consisting of two quadratic branches separated by the energy associated
with the dimer bond, $\gamma_1$, and the lower conduction band and upper
valence band are degenerate at the $K$ points of the Brillouin zone.
The existence of chiral charge carriers with a Berry's phase of
$2\pi$ was confirmed in the observation of the integer quantum Hall
effect \cite{novoselov:natphys2} where the low energy Landau level (LL)
spectrum is approximately linear in the field with $E_n \simeq \pm
\hbar\omega_c \sqrt{n(n-1)}$ for $n\geq0$ where $\omega_c$ is the
cyclotron frequency, and the spectrum includes a doubly degenerate LL at
zero energy \cite{mccann:prl96}.
It has been predicted theoretically \cite{mccann:prl96, mccann:prb74,
pereira:prb76} and observed experimentally
\cite{ohta:science313,castro:prl99}, that a gap can be induced in the
low energy band structure by breaking the symmetry between the
two layers. Switching of the conduction current by sweeping the Fermi
energy through the gapped region has been observed at low temperatures
\cite{oostinga:natmat7}, and this has lead to a surge of interest in
gapped BLG.

While the single particle theory of BLG is well known
\cite{mccann:ssc143, mccann:prl96, mccann:prb74, pereira:prb76}, it has
been shown that the electron-electron interactions are significant in
monolayer graphene \cite{iyengar:prb75,chakraborty:epl80}.  The Coulomb
interaction (CI) has been studied in the ungapped bilayer
\cite{nilsson:prb73}, while the biasing potential was considered in the
context of a ferromagnetic transition due to short-range interactions in
the mean-field approximation at zero magnetic field
\cite{castro:prl100},
and the absence of contribution to the intra-LL cyclotron
resonance from the electron interactions has been predicted within
Hartree-Fock theory \cite{barlas:prl101}.
However, the effect of the long range CI on the ground state of the
biased system in a magnetic field has not been investigated, and we
address this problem in the current Letter.

We find that the long range nature of the CI makes
significant changes to the properties of the low energy charge carriers
for BLG in a magnetic field. 
The interactions are significantly stronger for electrons
in the lowest LL, and this manifests itself in an observable way by
lifting the degeneracy of the cyclotron resonance transitions at filling
factors $\pm4$ \cite{henriksen:prl100,abergel:prb75}.
It also allows the possibility of mixing of the LLs which were
well-separated in energy when the CIs were not considered.  By
calculating the explicit form of the ground state wave function, we show
how this mixing fundamentally changes the nature of the ground state in
the biased bilayer, by inducing a finite spin polarization for
half-filled LLs.

\begin{figure}
	\centering
	\includegraphics{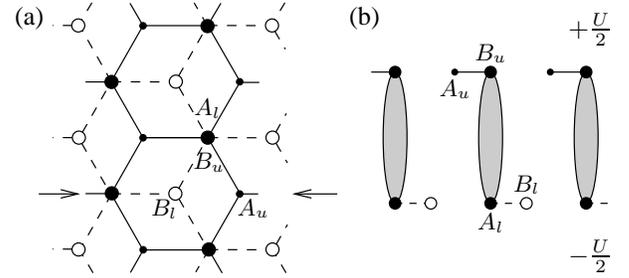}
	\caption{The lattice structure of bilayer graphene. The upper
	(lower) lattice is shown by solid (dashed) lines. 
	(a) The top-down view; (b) the side-on view projected along the axis
	between the two arrows in (a).
	\label{fig:lattice}}
\end{figure}

We model BLG as two sheets, each containing two inequivalent triangular
sublattices (labelled $A$ and $B$) of carbon atoms.  In the Bernal
stacking arrangement, the inter-layer bonds consist of dimers formed
from atomic orbitals associated with the $A$ sublattice in one layer and
the $B$ sublattice in the other (see Fig.  \ref{fig:lattice}), and the
energy associated with this bond is denoted $\gamma_1$ \cite{gamma3}.
We allow for a static electric potential $U$ to be applied between the
layers, so that the upper (lower) layer has potential $U/2$ ($-U/2$).
In a strong magnetic field we can write the tight-binding Hamiltonian
using a four component single valley basis where $\xi=\pm1$ labels the
valley \cite{mccann:prl96} as
\begin{equation}
	\mathcal{H}_0 = \begin{pmatrix}
		\frac{\xi U}{2} & 0 & 0 & \xi v_F \pi^\dagger \\
		0 & -\frac{\xi U}{2} & \xi v_F \pi & 0 \\
		0 & \xi v_F \pi^\dagger & -\frac{\xi U}{2} & \gamma_1 \\
		\xi v_F \pi & 0 & \gamma_1 & \frac{\xi U}{2} \end{pmatrix}
	\label{eq:Ham}
\end{equation}
where $\pi$ and $\pi^\dagger$ are the operators corresponding to
electron hops between neighboring atoms (in opposite sublattices in the
same layer).
The spectrum $\varepsilon_n^\xi$ is found from the quartic polynomial
\cite{pereira:prb76} derived from the Schr\"odinger equation associated
with the Hamiltonian in Eq. \eqref{eq:Ham}, where $n\in
\{0,1,2,\ldots\}$.  Additionally, we denote the band of a particular LL
by placing a `$+$' (`$-$') after the level index for the conduction
(valence) band. The wave functions associated with the LLs from the two
low energy bands are given by
\begin{equation}
	\psi_{n\pm}^\xi = e^{iky} \left( a_{n\pm}^\xi \varphi_{n+1}, \,
	b_{n\pm}^\xi \varphi_{n-1}, \,
	c_{n\pm}^\xi \varphi_n, \,
	d_{n\pm}^\xi \varphi_n \right)^T
	\label{eq:evec}
\end{equation}
where the functions $\varphi_n$ are the magnetic oscillator functions in
the Landau gauge, and the coefficients $a$, $b$, $c$, and $d$ are
defined so that the overall wave function is normalized to unity. 
There are also levels with $n=0\pm$, which have wave functions 
$\psi_{0+}^K = e^{iky} \left( \varphi_0, 0, 0, 0 \right)$, and
$\psi_{0-}^{K'} = e^{iky} \left( \varphi_0, 0, 0, 0 \right)$ with
$\varepsilon=\pm\delta$; and $\psi_{0+}^{K'}$ and $\psi_{0-}^K$ are
defined as higher LLs above with the appropriate substitutions for $n$
and $\xi$. When $U=0$, these four states are degenerate yielding the
eight-fold degeneracy (including the factor of 2 for spin) seen in the
integer quantum Hall effect in BLG \cite{novoselov:natphys2}.  
We include the fermionic properties of the electrons by constructing
Slater determinants for the non-interacting many body basis wave
functions.

To include the effects of the long-range CI we consider
the Hamiltonian
\begin{equation}
	\mathcal{H}_{\mathrm{Coul}} = \frac{1}{2} \sum_{i\neq j}
	\frac{e^2}{\epsilon |\vec{r}_i-\vec{r}_j|}
\end{equation}
where the vectors $\vec{r}_{i,j}$ label the positions of the electrons,
and $\epsilon=4\pi\epsilon_0\kappa$ is the dielectric constant of
graphene.  For graphene mounted on an SiO$_2$ substrate, $\kappa = 2.5$
\cite{graphene-dielectric}.
Our analysis is conducted by employing the exact
diagonalization scheme \cite{exactdiag} in which we calculate the
linear combination of non-interacting basis states which gives the
ground state of the Hamiltonian $\mathcal{H} = \mathcal{H}_0 +
\mathcal{H}_\mathrm{Coul}$. 
This method entails dividing the infinite sheet into rectangular cells
of dimension $L_x\times L_y$ \cite{asprat} and applying periodic
boundary conditions to the wavefunctions at the edges of each cell.
The matrix elements of the interaction over the single particle states
are evaluated exactly, and the interaction between the cells is taken
into account by adding the Madelung energy of a charged lattice
\cite{madelung}. 

The single particle states included in the Hilbert space from which the
non-interacting many body basis is constructed are as follows. 
There are four relevant quantum numbers: The LL index $n$, the
valley $\xi$, the spin and the momentum $k=\pi m/L_x$. 
The values of the momentum are fixed when the boundary conditions are
applied to the cell, and are labelled by $0\leq m \leq M-1$ with $M=L_x
L_y/(2\pi\lambda_B^2)$.
The LLs selected are governed by the details of the system we wish to
model, and $M$ is set by computational restraints.
Our model includes all inter-electron screening effects since we
calculate the exact matrix elements of the full Coulomb interaction, and
it is well known that filled Landau levels with energy
significantly below the Fermi level do not make additional
contributions.

In order to reduce the size of the many body system (and so improve
the calculation speed), we see that the Hamiltonian conserves the
total momentum $\mu=\sum_{i=1}^N m_i \mod M$. 
Therefore, we can perform a seperate diagonalization for each value of
$\mu$, and reduce the basis size to approximately the 1/$M$th part. 
We define $S=\sum_i S_i$ to be the total spin of the many electron
system, and since there is no spin-dependent term in the Hamiltonian,
$S_z$ (the projection of $S$ on the $z$ axis) is a good quantum
number. 
Therefore we fix $S_z$ to its minimum value whilst still being able to
recover all eigenstates of $S^2$ \cite{chakraborty:epl80}, further
reducing the many body basis size.

We consider two cases:
Firstly we demonstrate the strength of the interaction by calculating
the shift in the energy of each LL due to interactions for $U=0$
(an unbiased bilayer) and apply the results to recent experimental data.
Then we examine the system where the filling factor is negative, the
inter-layer potential sizeable, and the magnetic field strong. In this
case, we observe changes in the total spin of the ground state as a
function of $U$ and the magnetic field strength $B$.

\begin{table}[tb]
	\centering
	\setlength{\extrarowheight}{2pt}
	\newlength{\trind}
	\setlength{\trind}{0.52cm}
	\begin{tabular}{|p{2.8cm}|| >$c<$ | >$c<$ | >$c<$ | >$c<$ |}
	\hline
	(a) \hspace{0.03cm}Filling factor& -3 &    -2   &    -1   &    0    \\
	\hspace{\trind}Energy shift & -0.6443 & -0.6443 & -0.6443 & -0.6443 \\ 
	\hline
	\hspace{\trind}Filling factor&    1   &     2   &     3   &    4    \\
	\hspace{\trind}Energy shift & -0.6316 & -0.6222 & -0.6148 & -0.6085 \\ 
	\hline
	\end{tabular}
	\begin{tabular}{ |p{2.8cm}| |>$c<$ |>$c<$ |>$c<$ |>$c<$ |}
	\hline
	(b) Landau level $n$ &    1+   &    2+   &    3+   &    4+   \\ \hline
	\hspace{\trind}$\nu_n = 1$ & -0.4766 & -0.5001 & -0.5242 & -0.5160 \\
	\hspace{\trind}$\nu_n = 2$ & -0.4705 & -0.4880 & -0.5176 & -0.5110 \\
	\hspace{\trind}$\nu_n = 3$ & -0.4645 & -0.4759 & -0.5111 & -0.5061 \\
	\hspace{\trind}$\nu_n = 4$ & -0.4584 & -0.4638 & -0.5046 & -0.5012 \\ 
	\hline
	\end{tabular}
	\caption{Energy shift per electron due to the Coulomb interaction
	for integer filling factors in the (a) $0\pm$ LL and (b) $|n|\geq1$
	LLs, for $U=0$. 
	Energy units are $e^2/(\epsilon\lambda_B)$,
	the number of momentum states $M=3$, 
	and the magnetic field $B=3\unit{T}$.
	\label{tab:coulshift}}
\end{table}

We model the unbiased bilayer near half-filling by taking a single
particle Hilbert space consisting of electrons in the $0+$ and $0-$ LLs
with all possible spin and valley states at $U=0$. Each integer value of
the filling factor $\nu$ is
simulated by taking the number of electrons $N=(\nu+4)M$, and we have
$M=3$.  Table \ref{tab:coulshift}(a) shows the results of diagonalizing
the resulting many body Hamiltonian and evaluating the change in energy
from the non-interacting ground state for integer filling factors.  We
notice that the energy shift per electron reduces slightly as the LL
fills.

In Table \ref{tab:coulshift}(b), we show the energy shift due to
the CI for electrons in higher LLs
(\textit{i.e.} for levels with $|n|\geq 1$). 
We have taken a single particle Hilbert space consisting of all spin and
valley states within one LL.
The filling factor $\nu_n$ within LL $n$ can range between $0$
(corresponding to an empty level) and $4$ (a filled level), so that
$\nu_n=4$ and $\nu_{n+1}=0$ describe the same overall filling factor. 
The number of electrons is set by $N=\nu_n M$, and in order to allow
direct comparison with the lowest LL we restrict ourselves to $M=3$.
The energy associated with the interaction of electrons is very similar
in each of the higher LLs, and that the interaction energy per particle
is slightly reduced as the LL is filled.
We have verified that the results are identical in the valence band.

\begin{figure}[tb]
	\centering
	\includegraphics{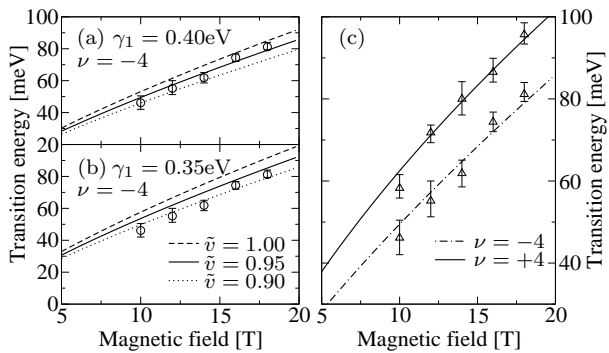}
	\caption{Electron-hole asymmetry in the inter-LL optical
	transition energy. The experimental data (represented as points) are
	taken from Ref. \onlinecite{henriksen:prl100}, Fig. 2. In (a) and
	(b), $\tilde{v}=v_F/(10^6\unit{ms^{-1}})$; in (c) we take
	$\gamma_1=0.4\unit{eV}$, $v_F=0.95\times10^6\unit{ms^{-1}}$;
	$M=6$ throughout.
	\label{fig:optics}}
\end{figure}

Together, Tables \ref{tab:coulshift}(a,b) show that the effect of the
long range CIs is considerable, and that for this value of the magnetic
field ($B=3\unit{T}$) the shift in the higher LLs is only about
two-thirds that of the lowest LL. 
This difference in the shift will reveal itself in the infra-red
absorption spectrum of bilayer graphene, since the energy of the optical
transitions depends entirely on the direct energy spacing between
levels.  At $U=0$ and $\nu=-4$, the lowest energy transition is from the
$1-$ level to the $0\pm$ level, while at $\nu=+4$ the lowest energy
transition is from $0\pm$ to $1+$. Therefore, if the $0\pm$ is shifted
with respect to the two $|n|=1$ levels, the degeneracy of these two
transitions predicted in the single particle theory \cite{abergel:prb75}
will be lifted. 
In Figure \ref{fig:optics} we show the predictions of our theory in
comparison to recent experimental data \cite{henriksen:prl100}. Panes
(a) and (b) show comparison of the valence band transitions
(\textit{i.e.} for $\nu=-4$) for two values of $\gamma_1$ and three
values of $v_F$.  The energy of the excited state is calculated as the sum
of the full interacting energy of $N-1$ electrons in the $1-$ level
and 1 electron in the $0\pm$ level (where we assume no LL mixing because
the magnetic field is strong). 
The transition energy is the difference between this and the energy of
$N$ interacting electrons in the $1-$ level. Pane (c) shows the
comparison of theoretical and experimental data in both bands for the
best values of parameters. The correspondence to our theory is clear.

\begin{figure}[bt]
	\centering
	\includegraphics[]{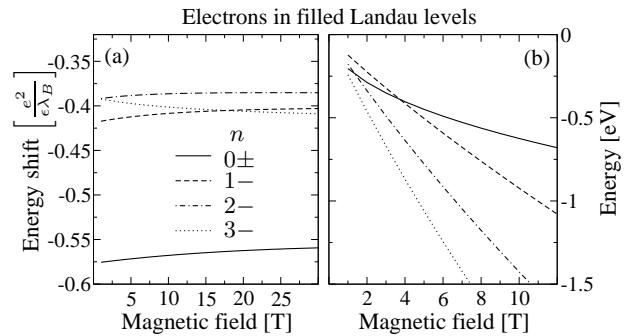}
	\caption{(a) The energy shift per electron of filled LLs.
	(b) The absolute energy per electron of filled LLs
	showing the crossing between the $n=0\pm$ degenerate level
	and the higher LLs in the valence band.
	In both plots $U=0$, and $M=5$.
	\label{fig:flplots}}
\end{figure}

In Fig. \ref{fig:flplots} we show the energy shift and absolute energy
of filled LLs as a function of the magnetic field. The
strength of the interaction scales with $e^2/(\epsilon\lambda_B) \propto
\sqrt{B}$ with a roughly constant coefficient, while the LL spacing goes
as $\hbar\omega_c \propto B$, so for lower values of the field, the
$n=0\pm$ level crosses the $2-$ and $1-$ levels as shown in Fig.
\ref{fig:flplots}(b).
The data shown here were calculated with $\kappa=2.5$, modelling
graphene \cite{graphene-dielectric} deposited on an SiO$_2$ substrate.
For suspended graphene (where $\kappa\approx 1$), it is conceivable
that the effect of the interaction would be even stronger.
Additionally, the effect of the inter-layer potential is to bring
together the valence band LLs with low index \cite{pereira:prb76}, so it
is plausible that the interactions will cause significant mixing between
these levels.

\begin{figure}[tb]
	\centering
	\includegraphics[]{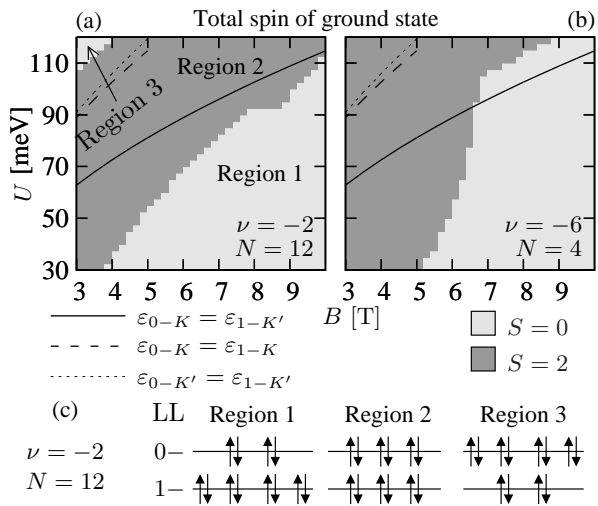}
	\caption{The total spin of the ground state of the (a) $\nu=-2$ and
	(b) $\nu=-6$ systems. $M=2$ and $N=(\nu+8)M$. 
	The lines show the crossing points of the single particle states.
	The graining is due to the finite interval between data	points.
	(c) The occupancy of the single electron states in the interacting
	many body ground state for each region of the plot in (a).
	The $z$-projection of the total spin is fixed at zero as described
	in the text.  \label{fig:spins}}
\end{figure}

Now we turn our attention to the system with an inter-layer potential
applied, so that $U\neq 0$. 
With a finite gap between the $0+$ and $0-$ levels and non-zero
filling factor, it is possible to consider the negatively-doped system
by taking only those single particle states which are in the valence
band. Therefore we select the single particle states which form the
Slater determinants describing the non-interacting basis states by
taking all spin and valley states of the $0-$ and $1-$ LLs.
We have $M=2$ and the number of electrons is related to the
filling factor by $N=(\nu+8)M$.  Diagonalizing this system for
half-filled LLs (so for $\nu=-6$ and $\nu=-2$), and
calculating the expectation value of the total spin operator $S^2$ over
the resulting ground state as a function of the magnetic field and the
inter-layer bias gives the data shown in the plots in Fig.
\ref{fig:spins}(a,b).
We have superimposed lines which represent the energy at which the
single particle energy levels cross, as labelled in the legend.
We have also calculated the expectation of the total valley quantum
number for each of these systems and find that it is constant with a
value equal to 2.
This is expected for half-filled LLs because of the lifting of the
valley degeneracy in the single particle theory by the inter-layer
potential \cite{mccann:prb74, pereira:prb76}.
The plots show that there is an abrupt change in the total spin of the
ground state, and a range of parameters where there is a non-zero
polarization of the spin. 
This transition is not directly related to the crossings of the single
particle states, since the position and slope of the transitions do not
match the corresponding lines superimposed on the plots.
This effect is therefore purely due to the CI, and in
particular to the exchange contribution which acts to minimize the
energy of spin-polarized many body states.

Figure \ref{fig:spins}(c) shows the occupation of the single particle
levels in the interacting many body ground state of the $\nu=-2$ system.
For simplicity, we display only the LL index of the states.
The actual ground state is a coherent combination of several 
non-interacting basis states, where the combination of LL
indices is consistent but different arrangements of momentum and valley 
states each come with identical prefactors in the linear combination.
In the lower-right region of the parameter space, the electrons occupy
as many of the $1-$ states as possible. Moving toward
the upper-left region, the $0-$ levels become successively more
populated. The absence of spin polarization in regions 1 and 3 is caused
by the pairing of electrons in the same valley. In region two, where
there are six electrons per LL, this pairing is incomplete and the spin
polarization finite.
The pattern of filling in the two regions of the $\nu=-6$ system is
identical.

In conclusion, we have shown that the long range CI between electrons
plays an important r\^ole near the Dirac point in BLG. In the unbiased
case, the interactions will cause a change in the cyclotron resonance
energies associated with the $0\pm$ LL, and LL mixing between
the $0\pm$ and $1-$ levels is induced for moderate magnetic fields.  
If an inter-layer bias is applied to split the valence and conduction
bands, the electron-electron interactions precipitate a transition in
the total spin of the ground state of half-filled LLs for certain ranges
of parameters.  Various experimental techniques to measure ground state
spin polarizations in the quantum Hall effect regime were reported
earlier \cite{spinexp} and might prove to be useful here as well.  This
effect will have fundamental implications for the design of devices made
from this material.

We acknowledge support from the Canada Research Chairs Program, and the
NSERC Discovery Grant.



\begin{thebibliography}{}
\bibitem{gamma3} We have not considered the effect of direct $A\to B$
	hops for sites in opposite layers, parameterized by $\gamma_3\approx
	0.3\unit{eV}$, since this does not lead to any additional states
	being coupled by the CI, and will therefore not make any qualitative
	changes to the physics we describe.
\end{thebibliography}
\end{document}